\documentclass[conference]{IEEEtran}
\IEEEoverridecommandlockouts
\usepackage{cite}
\usepackage{amsmath,amssymb,amsfonts}
\usepackage{algorithmic}
\usepackage{graphicx}
\usepackage{textcomp}
\usepackage{xcolor}
\def\BibTeX{{\rm B\kern-.05em{\sc i\kern-.025em b}\kern-.08em
    T\kern-.1667em\lower.7ex\hbox{E}\kern-.125emX}}

\usepackage{booktabs}
\usepackage[ruled,linesnumbered]{algorithm2e}
\usepackage{makecell}
\usepackage{soul}
\usepackage[most]{tcolorbox}
\usepackage{pifont}
\usepackage{multirow}

\begin{document}

\title{ALLMod: Exploring \underline{A}rea-Efficiency of \underline{L}UT-based \underline{L}arge Number \underline{Mod}ular Reduction via Hybrid Workloads}

\author{
\IEEEauthorblockN{Fangxin Liu$^{1,2,\dagger}$, Haomin Li$^{1,2,\dagger}$, Zongwu Wang$^{1,2}$, Bo Zhang$^{3}$, Mingzhe Zhang$^{3}$, Shoumeng Yan$^{3}$,\\ Li Jiang$^{1,2,*}$, and Haibing Guan$^{1}$}
\IEEEauthorblockA{\textit{1. Shanghai Jiao Tong University, 2. Shanghai Qi Zhi Institute, 3. Ant Group}\\
\IEEEauthorblockA   {*Corresponding Author \,\,\,\,\,\,\,\{liufangxin,\,haominli,\,ljiang\_cs\}@sjtu.edu.cn}
}
\thanks{$\dag$ These authors contributed equally.
This work was partially supported by the National Key Research and Development Program of China (2024YFE0204300), National Natural Science Foundation of China (Grant No.62402311), and Natural Science Foundation of Shanghai (Grant No.24ZR1433700).
This work was also supported by Ant Group through CCF-Ant Research Fund (CCF-AFSG RF20240304).
}
}

\maketitle

\begin{abstract}
Modular arithmetic, particularly modular reduction, is widely used in cryptographic applications such as homomorphic encryption (HE) and zero-knowledge proofs (ZKP). High-bit-width operations are crucial for enhancing security; however, they are computationally intensive due to the large number of modular operations required. The lookup-table-based (LUT-based) approach, a ``space-for-time'' technique, reduces computational load by segmenting the input number into smaller bit groups, pre-computing modular reduction results for each segment, and storing these results in LUTs. While effective, this method incurs significant hardware overhead due to extensive LUT usage.
In this paper, we introduce ALLMod, a novel approach that improves the area efficiency of LUT-based large-number modular reduction by employing hybrid workloads. Inspired by the iterative method, ALLMod splits the bit groups into two distinct workloads, achieving lower area costs without compromising throughput. We first develop a template to facilitate workload splitting and ensure balanced distribution. Then, we conduct design space exploration to evaluate the optimal timing for fusing workload results, enabling us to identify the most efficient design under specific constraints. Extensive evaluations show that ALLMod achieves up to $1.65\times$ and $3\times$ improvements in area efficiency over conventional LUT-based methods for bit-widths of $128$ and $8,192$, respectively.
\end{abstract}

% \begin{IEEEkeywords}
% Large Number Modular Reduction, Lookup Table, Privacy Computing
% \end{IEEEkeywords}

\section{Introduction}

Privacy computing, a crucial approach for safeguarding data security, has gained considerable attention in recent years due to increasing concerns about privacy and the protection of personal data on the Internet. This field encompasses a variety of applications, including homomorphic encryption (HE)\cite{acar2018survey}, which protects the privacy of user data and models, and zero-knowledge proofs (ZKP)\cite{goldwasser2019knowledge,fiege1987zero}, which ensure user privacy during transactions. These applications rely on public-key cryptography algorithms, such as Elliptic Curve Cryptography (ECC)\cite{koblitz1987elliptic} and RSA\cite{rivest1978method}, which use modular reduction as a fundamental component to prevent overflow.

The modular reduction used in these algorithms often requires extremely high bit-widths. For instance, ECC demands a bit-width of at least $224$ bits~\cite{chen2023digital}, while RSA typically requires at least $1,024$ bits~\cite{bos2009security}. Moreover, for most cryptographic schemes, higher bit-widths correspond to enhanced security~\cite{schneier2007applied}. Such high-bit-width operations pose challenges for efficiently deploying these algorithms and their applications on hardware. Common algorithms, like Barrett~\cite{barrett1986implementing} and Montgomery~\cite{montgomery1985modular}, implement modular operations through multiplication. However, as bit-width increases, constructing high-performance multipliers to meet latency and throughput requirements becomes increasingly difficult.

Recently, lookup-table-based (LUT-based) methods~\cite{devlin2019blockchain, ozturk2020design} have been proposed to achieve low-latency and high-throughput large-number modular reduction. This approach divides input bits into segments and pre-computes possible modular reduction results for each segment. These results are organized into lookup tables (LUTs), which can be implemented using block RAMs (BRAMs) on FPGAs. For each segment, the corresponding modular result is retrieved from the LUTs and combined through an adder tree to yield the final modular reduction result.

\begin{figure}[tp]
\centering
\setlength{\abovecaptionskip}{1pt}
\setlength{\belowcaptionskip}{1pt}
\includegraphics[width=1.0 \linewidth]{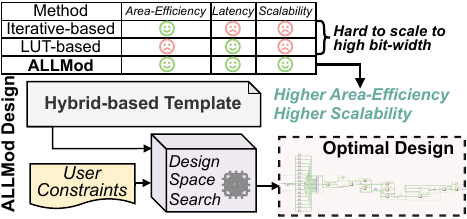}
\caption{\small Comparison of ALLMod to the existing modular reduction methods and ALLMod design overview.}
\label{fig: intro}
\vspace{-0.5cm}
\end{figure}

This approach converts complex modular reduction operations into simple lookup operations within LUTs, significantly reducing operational complexity and enabling high-throughput, low-latency large-number modular reduction. As a ``space-for-time'' trade-off, it utilizes BRAMs to store the LUTs and DSPs to implement adder trees that aggregate lookup results. However, as bit-width increases, this method faces the challenge of excessive resource overhead, resulting in suboptimal area efficiency.
On the other hand, the iterative method, though straightforward, requires only a minimal number of subtractors but exhibits high latency. 

To address these challenges, we propose combining iterative and LUT-based methods to enhance area efficiency in modular reduction implementations. Our solution, ALLMod, reduces area overhead by partitioning input bits into distinct workloads and distributing these across different modular modules, maintaining high throughput.
To facilitate workload partition and achieve balanced workload distribution, we design a template that guides implementation. Building on this, we introduce a design space exploration method to identify the optimal design under specified constraints. Experimental results show that ALLMod improves area efficiency by up to $3\times$ compared to the conventional LUT-based method.

In summary, our contributions are as follows:
\begin{itemize}
    \item We present a novel method to improve the area efficiency of LUT-based large-number modular reduction by smartly distributing the workload, enabling the integration of the iterative method with the LUT-based approach.
    \item We design a flexible template to facilitate workload division for implementations of varying bit-widths. This template, informed by latency analysis of both methods, ensures balanced workloads while minimizing area usage. 
    \item Using such a template, we propose a design space exploration approach focused on achieving high area efficiency. This approach includes a resource evaluator and an iterative search algorithm, enabling the identification of Pareto-optimal designs under specified constraints, such as latency and area requirements.
\end{itemize}

\section{Background}

\subsection{Large-Number Modular Reduction Methods}

In cryptographic algorithms, large-number modular reduction can be expressed as $R = A\ mod\ M$, where $A$ is a $2n$-bit number and $M$ is a fixed $n$-bit number.

\subsubsection{LUT-based Method}

The LUT-based method~\cite{devlin2019blockchain, ozturk2020design} employs lookup tables to store precomputed modular reduction results for possible segments of $A$, effectively reducing the computational load during runtime at the cost of additional space overhead. To leverage LUTs, the modular reduction operation must be transformed as follows:
\begin{equation}
    \label{eq: lut}
    \begin{aligned}
        &R = A\ mod\ M = \sum_{i=0}^{2n-1} (2^{i}a_{i})\ mod\ M\\
        &= (\sum_{i=n}^{2n-1} (2^{i}a_{i}) + \sum_{i=0}^{n-1} (2^{i}a_{i}))\ mod\ M\\
        &= ((\sum_{i=0}^{\frac{n}{k}-1} \sum_{j=0}^{k-1}(2^{n+ki+j}a_{n+ki+j})) + \sum_{i=0}^{n-1} (2^{i}a_{i}))\ mod\ M\\
        &= (\sum_{i=0}^{\frac{n}{k}-1} (\textcolor{red}{(\hat{a}_i\times 2^{n+ki})\ mod\ M}) + \textcolor{blue}{\sum_{i=0}^{n-1} (2^{i}a_{i})})\ mod\ M\\
    \end{aligned}
\end{equation}
where $\hat{a}_i = \sum_{j=0}^{k}(2^{j}a_{n+ki+j})$, i.e. $\hat{a}_i$ is a $k$-bit number.
The $\frac{n}{k}$ \textcolor{red}{red items} can be pre-computed offline and stored using LUTs, as the modulus $M$ is typically fixed.
Specifically, we use LUTs to pre-compute and store the values of $\hat{a}_i\times 2^{n+ki}\ mod\ M$.
A $k$-input and $n$-output LUT is used to store all the possible values of $\hat{a}_i\times 2^{n+ki}\ mod\ M$.
For the ($n=128$) case, an FPGA's BRAM with $36k$-bit capacity can be configured as an $8$-bit input and $128$-bit output LUT and $16$ BRAMs are required in total.
The retrieved results from LUTs and the \textcolor{blue}{blue item} are added together using a $\frac{n}{k}+1$-input adder tree to get an ($n+log_2d$)-bit number, where $d=\frac{n}{k}$.
The high $log_2d$ bits of the sum are then input to the LUTs for an additional lookup, and the result is added to the low $n$ bits. Finally, several subtractions are performed for adjustment.

\begin{algorithm}[tp]
    % \caption{LUT-based Modular Reduction~\cite{ozturk2020design}.}
    \caption{}
    \label{alg: lut}
    
    \let\oldnl\nl% Store \nl in \oldnl
    \newcommand{\nonl}{\renewcommand{\nl}{\let\nl\oldnl}}% Remove line number for one line
    
    \KwData{
    $2n$-bit number $A=\sum_{i=n}^{2n-1} (2^{i}a_{i})$, $n$-bit modulus $M$, and LUT's input bit-width $k$
    }
    \KwResult{$R\ =\ A\ mod\ M$}
    \BlankLine

    \nonl\textbf{Offline:}
    
    \For{$i\leftarrow 0,1,\dots,\lceil n/k \rceil-1$}
    {
        \For{$\hat{a}_i\leftarrow 0,1,\dots,2^k-1$}
        {
            \colorbox[RGB]{242, 242, 242}{
                \makebox[0.78\linewidth][l]{
                    \hspace{-0.4cm}
                    LUT[$i$][$\hat{a}_i$] $\leftarrow (\hat{a}_i\times 2^{n+ki})\ mod\ M$\ \textbf{Precompute}
                }
            }
        }
    }
    
    \BlankLine
    
    \nonl\textbf{Online:}
    
    $results\leftarrow$[] 

    \For{$i\leftarrow 0,1,\dots,\lceil n/k \rceil-1$ \textnormal{\textbf{parallel}}}
    {
        \colorbox[RGB]{226, 240, 217}{
            \makebox[0.84\linewidth][l]{
                \hspace{-0.4cm}
                $\hat{a}_i \leftarrow \sum_{j=0}^{k-1}(2^{j}a_{ki+j})$
            }
        } % \hfill \textbf{Segment High $\textit{\textbf{n}}$ Bits}

        \colorbox[RGB]{226, 240, 217}{
            \makebox[0.84\linewidth][l]{
                \hspace{-0.4cm}
                $results[i] \leftarrow$ LUT[$i$][$\hat{a}_i$]  \hfill \textbf{Parallel Look up}
            }
        }

    }

    \colorbox[RGB]{251, 229, 214}{
        \makebox[0.90\linewidth]{
            \hspace{-0.45cm}
            $R_0 \leftarrow$ \textbf{Add}($results$, $\sum_{i=0}^{n-1} (2^{i}a_{i})$) \hfill \textbf{Adder Tree}
        }
    }

    $results\leftarrow$[] 

    $n_r \leftarrow bitwidth(R_0)$

    \For{$i\leftarrow 0,1,\dots,\lceil (n_r-n)/k \rceil-1$ \textnormal{\textbf{parallel}}}
    {
        \colorbox[RGB]{226, 240, 217}{
            \makebox[0.84\linewidth][l]{
                \hspace{-0.4cm}
                $\hat{r}_i \leftarrow \sum_{j=0}^{k-1}(2^{j}R_0[{ki+j}])$ % \hfill \textbf{Segment High $\textit{\textbf{n}}$ Bits}
            }
        }

        \colorbox[RGB]{226, 240, 217}{
            \makebox[0.84\linewidth][l]{
                \hspace{-0.4cm}
                $results[i] \leftarrow$ LUT[$i$][$\hat{r}_i$]  \hfill \textbf{Parallel Look up}
            }
        }

    }

    \colorbox[RGB]{251, 229, 214}{
        \makebox[0.90\linewidth]{
            \hspace{-0.45cm}
            $R_1 \leftarrow$ \textbf{Add}($results$, $\sum_{i=0}^{n-1} (2^{i}R_0[{i}])$) \hfill \textbf{Adder (Tree)}
        }
    }

    \colorbox[RGB]{222, 235, 247}{
        \makebox[0.90\linewidth]{
            \hspace{-0.45cm}
            $R \leftarrow R_1\ or\ R_1-M\ or\ R_1-2M$\ \hfill \textbf{Subtract}
        }
    }
\end{algorithm}

Algorithm~\ref{alg: lut} outlines the detailed process of the LUT-based modular reduction.
The high $n$-bits are divided to $d$ segments, denoted as $\hat{a}_i$s, each containing $k$ bits.
Initially, each segment $\hat{a}_i$ is sent to the corresponding BRAM-based LUT to look up the pre-computed modular result of $\hat{a}_i\times 2^{n+ki}\ mod\ M$.
Then, the $d$ $n$-bit modular results, along with the low $n$-bit $\sum_{i=0}^{n-1} (2^{i}a_{i})$, are added together to obtain a $n+log_2d$-bit number $R_0$.
The high $log_2d$ bits of $R_0$ are then sent to the LUTs again, and the looked-up results are added with the low $n$ bits for $R_0$ to obtain $R_1$.
The second round of lookups helps refine $R_1$ to the final result $R$, assisted by several subtraction operations for adjustment.

\begin{algorithm}[tp]
    \caption{Iterative Modular Reduction~\cite{iterative}.}
    \label{alg: iterative}
    
    \let\oldnl\nl% Store \nl in \oldnl
    \newcommand{\nonl}{\renewcommand{\nl}{\let\nl\oldnl}}% Remove line number for one line
    
    \KwData{
    $2n$-bit number $A$, and $n$-bit modulus $M$
    }
    \KwResult{$R\ =\ A\ mod\ M$}
    \BlankLine
    
    $M\leftarrow M << n$\hfill \textbf{Align}

    \For{$i\leftarrow 1,2,\dots,n$}
    {
        \If{$A\geq M$}
        {
            \colorbox[RGB]{222, 235, 247}{
                \makebox[0.78\linewidth]{
                    \hspace{-0.4cm}
                    $A\leftarrow A-M$ \hfill \textbf{Subtract}
                }
            }
        }
        $M\leftarrow M >> 1$ \hfill \textbf{Shift}
    }
    $R\leftarrow A$
\end{algorithm}

\subsubsection{Iterative-based Method}

The iterative-based modular reduction method~\cite{iterative} is much more straightforward: the modular operation is performed by repeatedly subtracting the modulus from the input number until the result is smaller than the modulus.
To optimize the process, shift operations are employed during each iteration to facilitate efficient subtractions. It is worth noting that because the lower $n$-bits of the aligned modulus $M$ are zeros, the subtractor only needs to handle the subtraction of up to $n$-bit numbers.

\subsection{FPGA Implementation for Modular Reduction}

Modular reduction implementations are widely deployed on FPGAs due to their high-performance parallel computing capabilities and the flexibility they offer for designing custom solutions tailored to specific needs. These implementations are primarily divided into two categories: multiplier-based methods and LUT-based methods.

Notable examples of multiplier-based methods include Barrett’s algorithm~\cite{barrett1986implementing} and Montgomery’s algorithm~\cite{montgomery1985modular}, which rely on multiple high-performance high-bit-width multipliers to achieve high throughput and low latency. However, as bit-width increases, the complexity of designing these multipliers grows significantly, making it challenging to meet throughput requirements.

On the other hand, LUT-based methods~\cite{devlin2019blockchain, ozturk2020design}, as discussed earlier, are better suited to support high throughput in large-number modular reduction. However, with increasing bit-width, this approach faces significant resource overhead, resulting in poor area efficiency.
LUTMR~\cite{opasatian2023lookup} proposes using the FPGA’s native LUTs instead of BRAMs for implementing lookup tables, aiming to reduce latency and improve LUT usage by reusing patterns. While this optimization helps at the implementation level, it does not address the inherent limitations of the method itself, resulting in only modest improvements (no more than 5\% LUT savings).

\section{Analysis and Motivation}
\begin{figure}[tp] 
\centering
\setlength{\abovecaptionskip}{3pt}
\setlength{\belowcaptionskip}{3pt}
\includegraphics[width=1.0 \linewidth]{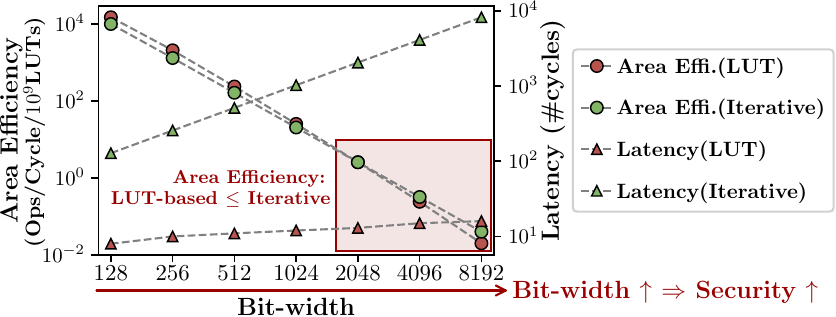}
\caption{\small Comparison of area efficiency and latency between LUT-based method and iterative-based method.}
\label{fig: profiling}
\vspace{-0.2cm}
\end{figure}

We conduct an in-depth analysis of the LUT-based and iterative methods for large-number modular reduction. Based on our profiling, we highlight key insights as follows:

\textbf{Observation: Trade-offs between area-efficiency and latency.}
We analyze the area efficiency and latency of both methods across various bit-widths, as shown in Figure~\ref{fig: profiling}.\footnote{ The overhead of BRAMs, adders, and subtractors is converted into LUTs, based on the utilization report from Vivado's IP generator.}
The LUT-based method offers lower latency but suffers from higher area overhead, while the iterative method exhibits better area efficiency at the cost of higher latency, especially for large bit-widths.
In the LUT-based approach, lookup tables reduce online computation, thus minimizing latency. However, the area overhead increases significantly with larger bit-widths, as more BRAMs are needed to store the LUTs, and larger adder trees are required for result aggregation. Moreover, as BRAM capacity is fixed, a higher bit-width limits the number of mappings a BRAM can store, increasing the number of result aggregations needed.

For example, with $n=128$, a $36k$-bit BRAM can store an $8$-bit input and $128$-bit output LUT, requiring 16 BRAMs and a 16-input, 128-bit adder tree. However, for $n=2,048$, a BRAM can only store a $4$-bit input and $2,048$-bit output LUT, requiring 512 BRAMs and an adder tree with 512 inputs, where each adder must support at least $2,048$-bit addition.

\textbf{Key Insight: Achieving Pareto Optimal Design through Method Fusion.} From the previous observation, the LUT-based method offers low latency but suffers from high area overhead, while the iterative method provides better area efficiency at the cost of higher latency. Therefore, it is feasible to balance area efficiency and latency by distributing the workload in a way that explores the Pareto optimal design space.
Existing methods have inherent trade-offs between area efficiency and latency, with no systematic approach to achieving both high area efficiency and low latency. This gap motivates our investigation into fusing the two methods to optimize modular reduction.

By leveraging the unique advantages of each method, we aim to reduce area consumption while maintaining efficient large-number modular reduction. In this paper, we analyze hybrid workloads across both methods to achieve enhanced performance and use this analysis to design a template for balanced workload distribution. We also explore the design space, considering area overhead and latency as flexible constraints, and identify the Pareto optimal boundary of our design. Finally, we propose an automatic search method to efficiently find the optimal design that aligns with user-specified constraints.

\section{Proposed ALLMod Design}

\subsection{Method Fusion based on Hybrid Workloads}

For formal modeling, we assume that a BRAM can store mappings from a $k$-bit input to an $n$-bit output.
We use $TP$, the number of modular reductions per cycle, to indicate the throughput of the design.
The upper bound of $TP$ is $MaxTP=0.5 Ops/cycle$, because each modular reduction requires $2$ cycles for the lookup in the BRAMs.
For a $2n$-bit input $A$, we divide it into two parts: the high $n-m$ bits are processed using the LUT-based method, while the low $n+m$ bits are handled by the iterative-based method. Here, $m$ is a design parameter that governs the workload distribution between the two methods.
The specific derivation of this partitioning is as follows:
\begin{equation}
    \label{eq: split}
    \begin{aligned}
        A &= \sum_{i=0}^{2n-1} (2^{i}a_{i}) = \sum_{i=n+m}^{2n-1} (2^{i}a_{i}) + \sum_{i=0}^{n+m-1} (2^{i}a_{i}) \\
        &= (\sum_{i=0}^{\frac{n-m}{k}-1}\sum_{j=0}^{k-1}(2^{n+m+ki+j}a_{n+m+ki+j}))+\sum_{i=0}^{n+m-1} (2^{i}a_{i}) \\
        &= \sum_{i=0}^{\frac{n-m}{k}-1} (\hat{a}_i\times 2^{n+m+ki}) + \sum_{i=0}^{n+m-1} (2^{i}a_{i}) \\
    \end{aligned}
\end{equation}
where $\hat{a}_i = \sum_{j=0}^{k-1}(2^{j}a_{n+m+ik+j})$.
Thus, the modular reduction can also be split into two parts of workloads:
\begin{equation}
    \begin{aligned}
        \underbrace{(\sum_{i=0}^{\frac{n-m}{k}-1} (\hat{a}_i\times 2^{n+m+ki}))\ mod\ M}_\textrm{\textbf{Workload I} for LUT-based method} + \underbrace{(\sum_{i=0}^{n+m-1} (2^{i}a_{i}))\ mod\ M}_\textrm{\textbf{Workload II} for iterative method}
    \end{aligned}
\end{equation}
In ALLMod, \textbf{workload I} is solved by the LUT-based method, and \textbf{workload II} is solved by the iterative-based method.
For scalability and area efficiency considerations, we employ serial adders to implement accumulation instead of an adder tree.

\begin{figure}[tp] 
\centering
\setlength{\abovecaptionskip}{3pt}
\setlength{\belowcaptionskip}{3pt}
\includegraphics[width=1.0 \linewidth]{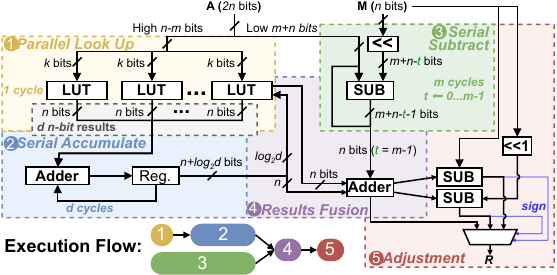}
\caption{\small ALLMod Template for balanced workload. Part \textcolor[RGB]{214, 182, 86}{\ding{172}} and part \textcolor[RGB]{108, 142, 191}{\ding{173}} support lookup and accumulation for LUT-based method. Part \textcolor[RGB]{130, 179, 102}{\ding{174}} supports serial subtraction for iterative-based method. Part \textcolor[RGB]{150, 115, 166}{\ding{175}} and part \textcolor[RGB]{184, 84, 80}{\ding{176}} are designed for fusing and adjusting the results.}
\label{fig: template}
\vspace{-0.3cm}
\end{figure}

\subsection{Template Design for balanced workload}
\label{sec: template}

Based on hybrid workloads, we propose a template design to maximize computation efficiency. As illustrated in Figure~\ref{fig: template}, the template consists of five parts.
Part \textcolor[RGB]{214, 182, 86}{\ding{172}} are $d$ BRAMs for storing the pre-computed results of $\hat{a}_i\times 2^{n+m+ki}\ mod\ M$, where $d=\frac{n-m}{k}$.
The high $n-m$ bits are fed into the BRAMs for parallel lookup.
Part \textcolor[RGB]{108, 142, 191}{\ding{173}} uses an adder for serially accumulating the $d$ results from the BRAMs.
Part \textcolor[RGB]{130, 179, 102}{\ding{174}} consists of a subtractor and a shifter for iterative subtractions.
Part \textcolor[RGB]{150, 115, 166}{\ding{175}} is an adder for fusing the results from the two methods, including the low $n$ bits of the accumulation results from part \textcolor[RGB]{108, 142, 191}{\ding{173}}, the second round lookup result from part \textcolor[RGB]{214, 182, 86}{\ding{172}}, and the serail subtraction result from part \textcolor[RGB]{130, 179, 102}{\ding{174}}.
Part \textcolor[RGB]{184, 84, 80}{\ding{176}} deploys several subtractors and a multiplexer for adjusting the final result.

The dataflow is also illustrated in Figure~\ref{fig: template}.
For the \textbf{left workload}, the high $n-m$ bits are sent to part \textcolor[RGB]{214, 182, 86}{\ding{172}} for parallel lookup, which takes one cycle. The results are then serially accumulated in part \textcolor[RGB]{108, 142, 191}{\ding{173}} over $d$ cycles. After accumulating the $n+\log_2d$-bit result in part \textcolor[RGB]{108, 142, 191}{\ding{173}}, the high $\log_2d$ bits are sent back to part \textcolor[RGB]{214, 182, 86}{\ding{172}} for a second round of lookup.
Meanwhile, for the \textbf{right workload}, the low $n+m$ bits are sent to part \textcolor[RGB]{130, 179, 102}{\ding{174}}, where they undergo $m-1$ cycles of subtraction.
In part \textcolor[RGB]{150, 115, 166}{\ding{175}}, the results from the previous three parts---part \textcolor[RGB]{214, 182, 86}{\ding{172}}, part \textcolor[RGB]{108, 142, 191}{\ding{173}}, and part \textcolor[RGB]{130, 179, 102}{\ding{174}}---are added together for fusion. The final result is then adjusted using several subtractions in part \textcolor[RGB]{184, 84, 80}{\ding{176}}.
Regarding resource reuse, since the execution of part \textcolor[RGB]{108, 142, 191}{\ding{173}} and part \textcolor[RGB]{130, 179, 102}{\ding{174}} does not overlap with that of part \textcolor[RGB]{150, 115, 166}{\ding{175}} and part \textcolor[RGB]{184, 84, 80}{\ding{176}}, resource reuse is possible. The adder in part \textcolor[RGB]{108, 142, 191}{\ding{173}} can be reused in part \textcolor[RGB]{150, 115, 166}{\ding{175}} for aggregation, and the subtractor in part \textcolor[RGB]{130, 179, 102}{\ding{174}} can also be reused in part \textcolor[RGB]{184, 84, 80}{\ding{176}} for adjustment. This further optimizes the design by reducing redundant hardware resources and enhancing efficiency.

Given that the two workloads are executed in parallel, the latencies of the two workloads should ideally be equal to ensure a balanced workload and minimize resource waste. The latency of the left workload is $d+1$ cycles ($d=\frac{n-m}{k}$), and the latency of the right workload is $m$ cycles. Therefore, the following equation should be satisfied:
\begin{equation}
    \begin{aligned}
        \frac{n-m}{k}+1=m\ \Rightarrow\ \ 
        m=\frac{n+k}{k+1}
    \end{aligned}
\end{equation}
Apart from the BRAMs, the adder and subtractor in the template need to be copied $d\times TP$ times to ensure the throughput.

The template can reduce the area overhead of both BRAMs and adders/subtractors.
Compared to the LUT-based method, the template can achieve $\frac{m}{n}=\frac{n+k}{n(k+1)}\approx \frac{1}{k+1}$ BRAM savings.
Moreover, the original LUT-based method needs $\frac{2n}{k}$ adders/subtractors, while ours just needs $d=\frac{n-m}{k}=$ pieces to reach maximum throughput $MaxTP$.
For $n=8,192$, a BRAM can only store mappings with $2$-bit input, meaning $k=2$.
Consequently, our template can save up to $50\%$ of the required BRAMs and over $65\%$ of the required adders/subtractors.

\subsection{Template-aided Design Space Exploration}

To accommodate a broader range of design requirements, we explore the design space of ALLMod. Using the proposed template, which focuses on balanced workloads, we investigate the design space by considering both latency and area constraints as key parameters.

\subsubsection{Parallel-Serial Hybrid for More Strict Latency Constraint}

In cases where stricter latency requirements are imposed, the single-pass algorithm must be completed in fewer cycles. Since the iterative method cannot be parallelized, we propose reducing its workload while shifting more work to the table lookup method. Additionally, we aim to enhance the parallel processing capacity of specific components within the table lookup method.

To achieve this, we introduce a small adder tree into part \textcolor[RGB]{108, 142, 191}{\ding{173}} of the template, which works alongside the serial accumulator to accelerate the accumulation process. This setup allows the workload in the left part of the design to be completed in fewer cycles, helping to meet the stricter latency constraints.
When configured with an adder tree that has $x$ inputs, part \textcolor[RGB]{108, 142, 191}{\ding{173}} can be completed in $\frac{n-m}{k}-x$ cycles, while part \textcolor[RGB]{130, 179, 102}{\ding{174}} will still take $m$ cycles. The optimal design point is determined by finding $m$ and $x$ that satisfy the following conditions:
\begin{equation}
    \begin{aligned}
        max(1+\frac{n-m}{k}-x, m)\leq Latency_{req}
    \end{aligned}
\end{equation}
where $Latency_{req}$ is the user-specified number of cycles required for the design.

\subsubsection{More Strict Area Constraint}

When area constraints become more stringent, we assign a larger portion of the workload to part \textcolor[RGB]{130, 179, 102}{\ding{174}} (iterative subtraction) to reduce the BRAM overhead of part \textcolor[RGB]{108, 142, 191}{\ding{173}} (parallel lookup). However, since the workload in part \textcolor[RGB]{130, 179, 102}{\ding{174}} cannot be accelerated in parallel, we need to ensure throughput by duplicating more of the components in part \textcolor[RGB]{130, 179, 102}{\ding{174}}.

Assuming that part \textcolor[RGB]{130, 179, 102}{\ding{174}} is duplicated $y$ times, its throughput is approximately $\frac{y}{m} \text{Ops/cycle}$. To guarantee throughput, we set $y = m \times TP$. Similarly, part \textcolor[RGB]{108, 142, 191}{\ding{173}} must be duplicated $d \times TP$ times. Consequently, the task is to determine the value of $m$ such that the design meets both area and throughput requirements:
\begin{equation}
    \begin{aligned}
        \frac{n-m}{k} \times &Area_{BRAM} + d\times TP \times Area_{Adder} \\ + m\times &TP \times Area_{Subtractor} \leq \text{Area}_{req}
    \end{aligned}
\end{equation}
where $Area_{req}$ is the user-specified area limit, and $Area_{BRAM/Adder/Subtractor}$ represents the area of a BRAM/adder/subtractor.

\begin{algorithm}[tp]
    \caption{Automatic Design Space Search.}
    \label{alg: search}
    
    \let\oldnl\nl% Store \nl in \oldnl
    \newcommand{\nonl}{\renewcommand{\nl}{\let\nl\oldnl}}% Remove line number for one line
    
    \KwData{
        Bit-width $n$, Input bit-width of BRAM(LUT) $k$, user-specified latency and area constraints $Latency_{req}$ and $Area_{req}$, expected throughput $TP$
    }
    \KwResult{Pareto optimal scheme list $OptSchemeList$}
    \BlankLine

    $SchemeList \leftarrow [\ ]$

    \For{$m \leftarrow 0 \dots n$}{
        % // n-m bits for LUT-based method, n+m bits for iterative method
        \For{$Width_{Tree}\leftarrow 0\dots n-m$}{
            $Latency_{LUT} \leftarrow 1 + max(\frac{n-m}{k}-Width_{Tree}$, $log_2(Width_{Tree}))$
            
            $Latency_{iter}\leftarrow m$

            $Latency$ $\leftarrow$ $max$ ($Latency_{LUT}$, $ Latency_{iter}$ )

            \If{$Latency \leq Latency_{req}$}{
                % // Latency requirement satisfied
                
                $N_{BRAMs}\leftarrow \frac{n-m}{k}$

                $N_{Adders_{Tree}}\leftarrow 2\times Width_{Tree}-1$

                $N_{Adders_{Serial}}\leftarrow (\frac{n-m}{k}-Width_{Tree}) \times TP$

                $N_{Adders} \leftarrow N_{Adder_{Tree}} + N_{Adder_{Serial}}$

                $N_{Subtractors} \leftarrow m\times TP$

                $Area\leftarrow$ \textbf{AreaEstimate}($N_{BRAMs}$, $N_{Adders}$, $N_{Subtractors}$)

                \If{$Area \leq Area_{req}$}{
                    $Scheme_{feasible} \leftarrow$ ($m$, $Width_{Tree}$, $Cycles$, $Area$)

                    % $Metrics \leftarrow$ ($Cycles$,$Area$)

                    SchemeList.append($Scheme_{feasible}$)
                    % $OpList\leftarrow$ \textbf{ParetoUpdate}($OpList$, $CandidateScheme$, $Metrics$)
                }
            }
        }
    }
    OptSchemeList $\leftarrow$ \textbf{FindParetoOptimal}($SchemeList$)
\end{algorithm}

\begin{table*}[thbp]
    \centering
\setlength{\abovecaptionskip}{3pt}
\setlength{\belowcaptionskip}{3pt}
    \caption{Comparison of LUT-based method, iterative method, and ALLMod (template), all aligned to $MaxTP$.}
    \resizebox{\linewidth}{!}{
    \begin{tabular}{|@{\hspace{2pt}}c@{\hspace{2pt}}||l@{\hspace{2pt}}|l@{\hspace{2pt}}|l@{\hspace{2pt}}||l@{\hspace{2pt}}|l@{\hspace{2pt}}|l@{\hspace{2pt}}||l@{\hspace{2pt}}|l@{\hspace{2pt}}|l@{\hspace{2pt}}||@{\hspace{2pt}}l@{\hspace{2pt}}|}
    \hline
    \multirow{2}{*}{Bit-Width ($n$)} &
      \multicolumn{3}{@{\hspace{2pt}}c@{\hspace{2pt}}||}{\textbf{Area Efficiency} (\textit{Ops/Cycles/$10^9$LUTs})} &
      \multicolumn{3}{@{\hspace{2pt}}c@{\hspace{2pt}}||}{\textbf{Latency} (\textit{Cycles})} &
      \multicolumn{3}{@{\hspace{2pt}}c@{\hspace{2pt}}||@{\hspace{2pt}}}{\textbf{Breakdown} (\textit{BRAMs;Adders;Subtractors})} &
      \textbf{Hybrid Workloads} \\ \cline{2-11} 
     &
      LUT-based~\cite{ozturk2020design} &
      Iterative~\cite{iterative} &
      \textbf{ALLMod} (Improve) &
      LUT-based &
      Iterative &
      \textbf{ALLMod} &
      LUT-based &
      Iterative &
      \textbf{ALLMod} &
      \textbf{ALLMod} \\ \hline
    128  & 15258.79 & 10172.53 & \textbf{25040.06} (1.65$\times$) & 9  & 128  & 20   & 16;3;1      & 0;0;64   & 15;8;8         & 113:143    \\ \hline
    256  & 2111.49  & 1326.85  & \textbf{4521.12} (2.14$\times$) & 11 & 256  & 37   & 37;73;1     & 0;0;128  & 32;16;16       & 224:288    \\ \hline
    512  & 241.60   & 165.86   & \textbf{550.18} (2.28$\times$) & 12 & 512  & 78   & 86;171;1    & 0;0;256  & 73;37;37       & 438:586    \\ \hline
    1024 & 25.75    & 20.73    & \textbf{61.05} (2.37$\times$) & 13 & 1024 & 176  & 205;409;1   & 0;0;512  & 171;86;86      & 853:1195   \\ \hline
    2048 & 2.59     & 2.59     & \textbf{6.45} (2.49$\times$) & 14 & 2048 & 415  & 512;1024;1  & 0;0;1024 & 410;205;205    & 1638:2458  \\ \hline
    4096 & 0.24     & 0.32     & \textbf{0.65} (2.71$\times$) & 16 & 4096 & 1029 & 1366;2731;1 & 0;0;2048 & 1024;512;512   & 3072:5120  \\ \hline
    8192 & 0.02     & 0.04     & \textbf{0.06} (\textbf{3.00$\times$}) & 17 & 8192 & 2736 & 4096;8191;1 & 0;0;4096 & 2731;1366;1366 & 5461:10923 \\ \hline
    \end{tabular}
    }
    \label{tab: overall}
    \vspace{-0.3cm}
\end{table*}

\begin{figure*}[tp] 
\centering
\setlength{\abovecaptionskip}{-1pt}
\setlength{\belowcaptionskip}{3pt}
\includegraphics[width=1 \linewidth]{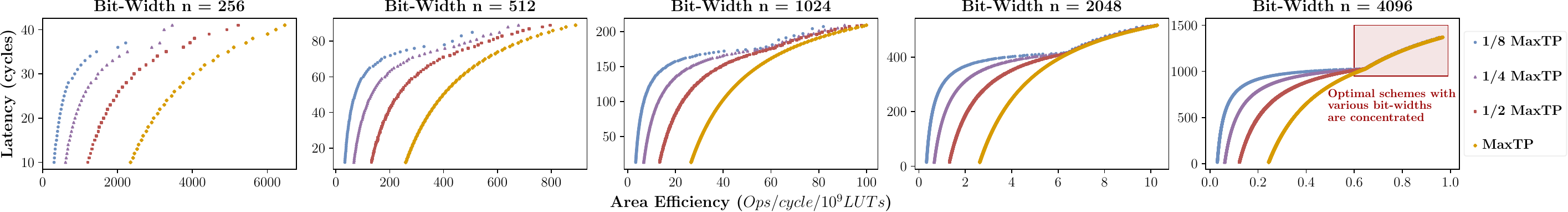}
\caption{\small Pareto optimal schemes of ALLMod for various throughputs and bit-widths. The optimal schemes with high area efficiency gradually converge as the bit-width increases.}
\label{fig: tp}
\vspace{-0.5cm}
\end{figure*}

\begin{figure}[tp] 
\centering
\setlength{\abovecaptionskip}{3pt}
\setlength{\belowcaptionskip}{3pt}
\includegraphics[width=1.0 \linewidth]{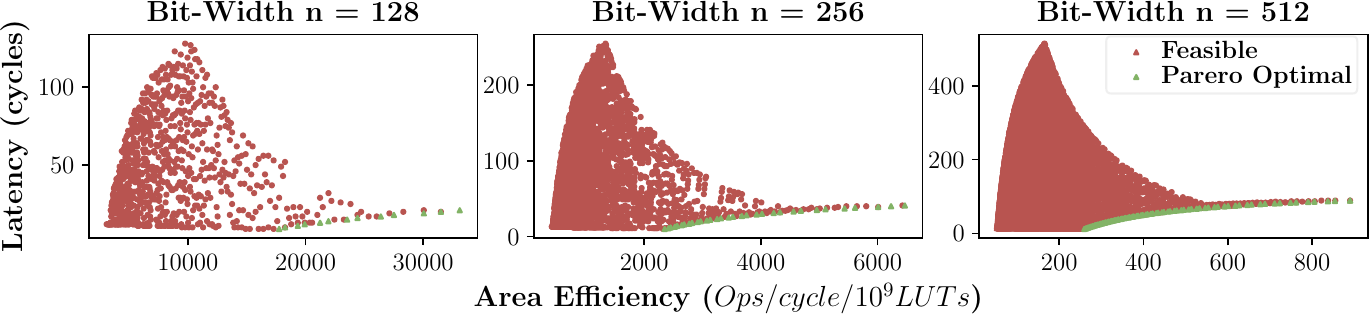}
\caption{\small Visualization of searched schemes for various bit-width at maximum throughput $MaxTP$. \textbf{\textcolor[RGB]{184, 84, 80}{Red}} points indicate the feasible schemes and \textbf{\textcolor[RGB]{130, 179, 102}{green}} points indicate the Pareto optimal schemes.}
\label{fig: dse}
\vspace{-0.4cm}
\end{figure}

\subsubsection{Automatic Search}

The analysis of the design space only focuses on either lower latency or smaller area consumption.
In practice, the design usually has constraints on both latency and area.
Therefore, based on the previous analysis, we propose an automatic search approach to find the Pareto optimal design schemes that meet various constraints provided by users.
As shown in Algorithm~\ref{alg: search}, our search method takes the bit-width $n$, and user-specified constraints as input, and outputs a list of Pareto optimal schemes.
Users' constraints include latency requirement $Latency_{req}$, area requirement $Area_{req}$, and expected throughput $TP$.

First, we enumerate $m$ from $0$ to $n$ to explore different workload segmentations. Specifically, $m=0$ corresponds to the pure LUT-based method, while $m=n$ corresponds to the pure iterative method. Next, we enumerate the input width of the adder tree, $Width_{Tree}$, from $0$ to $n-m$. When $Width_{Tree}=0$, the template with a balanced workload, as introduced in Section~\ref{sec: template}, is used; when $Width_{Tree}=n-m$, part \textcolor[RGB]{108, 142, 191}{\ding{173}} is entirely implemented using an adder tree for parallel accumulation. Each pair of values ($m$, $Width_{Tree}$) forms a key \textit{parameter pair} for a candidate scheme.

For each \textit{parameter pair}, we compute the cycles required to complete the modular reduction, as outlined in Section~\ref{sec: template}, and verify whether the latency requirement is met. Then, we calculate the number of BRAMs, adders, and subtractors needed for the scheme and estimate the corresponding area overhead. Feasible schemes that satisfy both latency and area constraints are added to a candidate list, $SchemeList$. Finally, we identify the Pareto optimal schemes from this list.
Through such an automatic search process, users can easily identify the Pareto optimal design schemes that meet their specific constraints.

In addition, our approach efficiently identifies Pareto optimal schemes. By enumerating both $m$ and $Width_{Tree}$, the total number of feasible schemes, denoted as $S$, can be at most $n^2$. Using a priority queue to sort and select the optimal schemes, the time complexity for finding the Pareto optimal solutions is $O(S \log S) = O(n^2 \log n)$. Even for very large values of $n$, such as $8,192$, the entire search process completes in approximately ten minutes\footnote{On a laptop with Intel i9-12900H CPU and 32GB memory.}.

\section{Evaluations}

\subsection{Evaluation Methodology}
We compare ALLMod with the standard LUT-based method~\cite{ozturk2020design} and iterative method~\cite{iterative}.
We implement a $128$-bit prototype system for ALLMod on FPGA using Verilog, with a working frequency set to 200MHz.
To validate the timing and functionality of our models, we synthesized them using the Xilinx Vivado Design Suite~\cite{vivado}.
For evaluation, we use $\frac{Ops/cycle}{10^9 LUTs}$ as the metric of area efficiency. 
Given that the working frequency for both the baseline and our method is primarily determined by the latency of the n-bit adder/subtractor, we utilize $cycles$ as a standardized metric for latency comparison, assuming a similar frequency across implementations. This allows for a fair comparison of execution time independent of absolute clock speeds. The overhead of BRAMs, adders, and subtractors is converted with LUTs, based on the utilization report from the IP generator of Vivado.

\subsection{Overall Performance}

Table~\ref{tab: overall} compares the performance of our ALLMod template with the baseline method. The methods are all aligned to the maximum throughput $MaxTP$ that the LUT-based method can achieve.
We show the area efficiency, latency, and the breakdown of resources for each method, and segmentation for hybrid workloads of ALLMod.

For small bit-width $n=128$, ALLMod achieves the area efficiency of $25,000\ Ops/Cycles/10^9LUTs$, which is $1.65\times$ higher than the LUT-based method.
For large number modular reduction with $n=8,192$, ALLMod achieves up to $3.0\times$ area efficiency improvement over the LUT-based method.
It's worth noting that although the LUT-based method has lower latency, it needs too much area overhead to achieve the same throughput as ALLMod.

The breakdown results demonstrate that ALLMod can save a significant number of BRAMs and adders/subtractors.
For $n=8,192$, ALLMod saves over $30\%$ of BRAMs and $66\%$ of adders/subtractors compared to the LUT-based method. 
As the bit-width increases, the proportion of workload allocated to the right parts in the template gradually increases, which illustrates the significance of introducing a hybrid workload design for optimizing the LUT-based method.

\subsection{Pareto Optimal with various throughput requirements}

Figure~\ref{fig: tp} shows the Pareto optimal schemes of ALLMod for various throughputs and bit-widths.
We iterate the throughput requirement from $MaxTP/8$ to $MaxTP$ for each bit-width.
As the bit-width increases, the Pareto optimal schemes for different throughputs are more concentrated in the region with high area efficiency, meaning that the design gradually converges.
Such a convergence demonstrates that ALLMod can effectively explore the design space and find optimal schemes for guiding the design.

\subsection{Design space exploration with automatic search}

Figure~\ref{fig: dse} visualizes the design space exploration results of ALLMod for various bit-widths.
The distributions of feasible schemes under different bit-widths are basically the same.
A significant concentration of feasible schemes is observed within the region of lower area efficiency and high latency, whereas a small number of feasible solutions are identified in the region of higher area efficiency.
This distribution underscores the criticality of identifying Pareto optimal schemes.
Our automatic search effectively filters out schemes with low area efficiency and high latency and finds area-efficient schemes with low latency.

\section{Conclusion}
This paper presents ALLMod, an area-efficient method for large-number modular reduction, combining LUT-based and iterative approaches. We then design a balanced workload template to guide the segmentation. Additionally, we present an automatic search approach to explore the design space of ALLMod based on this template. Evaluations show that ALLMod outperforms the LUT-based method in terms of area efficiency and latency.

\bibliographystyle{IEEEtran}
\bibliography{ref}

\end{document}